\documentclass[sigconf]{acmart}

\usepackage[utf8]{inputenc}
\usepackage{acronym}
\usepackage{color}
\usepackage{soul}
\usepackage{url}
\usepackage{amsmath}
\usepackage{graphicx}
\usepackage{balance}
\usepackage{bm}
\usepackage{amsthm}
\usepackage{todonotes}
\newtheorem{definition}{Definition}  
\usepackage{comment}

\usepackage[linesnumbered, ruled]{algorithm2e}
\usepackage{mwe}
\usepackage{pgf-umlsd}
\usepackage{tikz}
\usetikzlibrary{decorations.pathreplacing, positioning}
\tikzstyle{vertex}=[]

\SetKwRepeat{Do}{do}{while}

\acrodef{CI}{Critical Infrastructure}
\acrodef{DP}{Differential Privacy}
\acrodef{Geo-Ind}{Geo-Indistinguishability}
\acrodef{UAV}{Unmanned Aerial Vehicle}
\acrodef{LBS}{Location-Based Service}
\acrodef{TPR}{True Positive Ratio}
\acrodef{FPR}{False Positive Ratio}
\acrodef{PoI}{Point of Interest}
\acrodef{FAA}{Federal Avionics Administration}
\acrodef{AD}{\emph{Average Distance}}
\acrodef{TTP}{Trusted Third Party}
\acrodef{VANET}{Vehicular Ad-hoc Networks}

\setcopyright{acmcopyright}

\acmDOI{xx.xxx/xxx_x}

\acmISBN{978-1-4503-8713-2/22/04}

\acmConference[ARES'22]{17th International Conference on Availability, Reliability and Security}{August 23 – August 26, 2022}{Wien, Austria}
\acmYear{2022}
\copyrightyear{2022}
\acmArticle{4}
\acmPrice{15.00}

\begin{document}

\title[Hide and Seek: Privacy-Preserving and FAA-compliant Drones Location Tracing]{Hide and Seek: \\Privacy-Preserving and FAA-compliant Drones Location Tracing}

\author{Alessandro Brighente}
\email{alessandro.brighente@unipd.it}
\affiliation{%
  \institution{University of Padua}
  \city{Padua}
  \country{Italy}
}

\author{Mauro Conti}
\email{mauro.conti@unipd.it}
\affiliation{%
  \institution{University of Padua}
  \city{Padua}
  \country{Italy}
}

\author{Savio Sciancalepore}
\email{s.sciancalepore@tue.nl}
\affiliation{%
  \institution{Eindhoven University of Technology (TU/e), Department of Mathematics and Computer Science}
  \city{Eindhoven}
  \country{The Netherlands}
}

\begin{abstract}

Due to the frequent unauthorized invasions by commercial drones to Critical Infrastructures (CIs), the US-based Federal Avionics Administration (FAA) recently published a new specification, namely \emph{RemoteID}. Such a rule requires all drones to broadcast information about their identity and location, to allow for immediate invasion attribution and counter-actions. However, the enforcement of such a rule poses severe concerns on drones' operators, especially in terms of location privacy and tracking threats. Indeed, by simply receiving wireless signals, an adversary could know the precise drone location, track it, and infer sensitive information.

In this paper, we demonstrate that CI operators can detect timely and efficiently drones invading no-fly zones, without sacrificing drones location privacy. Specifically, we provide two major contributions. First, we propose DiPrID, the first RemoteID-compliant solution enhancing drones location privacy via Differential Privacy. Second, we introduce and evaluate ICARUS, an effective area invasion detection technique capable of identifying invasions by unauthorized drones, even when equipped with DiPrID, with remarkable accuracy and negligible detection delays.
Our experiments showed that when drones obfuscate their location by an average distance as large as $31.914$~km, ICARUS can detect up to $94.2$\% of invasions, while false-positives can be mitigated through the help of the FAA.
\end{abstract}



\maketitle

\section{Introduction}
\label{sec:intro}

\acp{UAV}, also known as \emph{drones}, are becoming increasingly popular. The proliferation of these new devices is mainly due to the plethora of application domains where they can provide concrete benefits, e.g., surveillance of \acp{CI}, amateur video shooting, and goods delivery~\cite{shakhatreh2019unmanned}. 

Unfortunately, \acp{UAV} can also be used for malicious purposes~\cite{sciancalepore2020pinch}. Indeed, the adoption of \acp{UAV} vastly simplifies unauthorized recordings of protected areas and the release of dangerous materials in the proximity of \acp{CI}~\cite{drones_saudi}. 
Only in 2017, 385 incidents involving small drones were reported~\cite{seedscience}.
To curb the increasing concerns, the US-based \ac{FAA} recently published a dedicated rule, namely \emph{RemoteID}, just become effective~\cite{remoteID}. As per this rule, all \acp{UAV}, mostly independently from their weight and usage, should broadcast wirelessly in clear-text information about their identity, location, speed, and emergency status. The availability of such information could enable the unique identification and location of unauthorized drones, simplifying the adoption of follow-up countermeasures and guilt attribution.
Complying with this rule will be mandatory from late 2022, and both UAV manufacturers and operators will be fined in case of non-compliance. 
 
Although the favourable reactions from the CI community, \emph{RemoteID} also generated concerns on \acp{UAV} operators~\cite{ainonline}. Indeed, the uncontrolled broadcast of identity and location information could enable several attacks against \acp{UAV}, including UAV tracking, capturing, flight disruption and privacy leakages. For instance, goods delivery companies such as Amazon would immediately reveal the location of their storage or the site of the customer requesting specific goods~\cite{dronedj_privacy}. At the same time, UAV operators also argue that neither the FAA nor the CI requires continuous knowledge of the UAV position to detect them, as systems to detect area invasions based on other means are already deployed and working with acceptable performances.

The objective of protecting \acp{UAV}' location privacy while not degrading data utility shares similarities with issues tackled in the context of \ac{LBS}. There, users would like to \emph{obfuscate} their location not to reveal their movement patterns to the servers. At the same time, obfuscating too much the reported location would make the released data useless. In that context, the issue was tackled by \emph{obfuscating} the location of the user through dedicated mechanisms, such as \ac{Geo-Ind}, borrowed from the well-known concept of \ac{DP}, traditionally applied for secure databases querying.
However, to the best of our knowledge, previous contributions did not investigate similar issues in the context of \acp{UAV}. Specifically, the existing literature misses a systematic investigation of the effects of the obfuscation of the \acp{UAV} location, on both their privacy and the capabilities of \ac{CI} operators to detect invasions of \emph{no-fly zones} timely. Also, there is no system currently able to detect invasions in such a scenario and obtain \acp{UAV} actual location.

{\bf Contribution.} In this paper, we investigate the trade-offs between location privacy and data utility in UAVs operations, in the framework of the \emph{RemoteID} rule. Specifically, we provide the following contributions. 
\begin{itemize}
    \item Through experimentation on real data, we demonstrate  that \ac{CI} operators can detect \acp{UAV} invading no-fly zones, without fully jeopardizing \acp{UAV} location privacy.
    \item We propose DiPrID, the first standard-compliant extension to \emph{Remote ID}, enabling differentially-private location disclosure and encrypted location reports, decryptable only by a \ac{TTP} (e.g., the FAA).
    \item We introduce ICARUS, a simple yet effective strategy allowing \ac{CI} operators to timely detect \emph{no-fly zones} invasions, even when invading objects integrate DiPrID.
    \item We demonstrate that ICARUS could detect invasions by UAVs equipped with DiPrID with an outstanding accuracy of $94.2\%$. At the same time, \acp{UAV} can maintain location privacy, broadcasting locations at an average distance of $31.914$~km, at the expense of a few false positives.
    \item We provide a framework enabling the \acp{UAV} and CI operators to discover false positives thanks to the FAA, being this latter the only one that can unveil the actual UAV location.
\end{itemize}

Our study, performed over a real dataset of \acp{UAV} flights, demonstrates that it is possible to find a balance between the degree of location privacy offered to the \acp{UAV} and the detection capabilities of the \ac{CI} monitoring stations. Such a trade-off comes at the expense of a small detection delay on the CI operator and a non-negligible probability of false invasion attributions, solvable by the TTP. 

{\bf Roadmap.} This paper is organized as follows. \mbox{Sec.~\ref{sec:background}} introduces the preliminaries, \mbox{Sec.~\ref{sec:scenario_adv_model}} depicts the scenario and adversarial model, \mbox{Sec.~\ref{sec:proposals}} describes DiPrID and ICARUS, \mbox{Sec.~\ref{sec:performance}} shows our results, \mbox{Sec.~\ref{sec:related}} outlines the related work, and finally, \mbox{Sec.~\ref{sec:conclusion}} tightens conclusions.

\section{Background}
\label{sec:background}

We first introduce the RemoteID specification in Sec.~\ref{sec:remoteid}, then summarize the notions of \ac{DP} and \ac{Geo-Ind} in Sec.~\ref{sec:diff_privacy}.

\subsection{The \emph{RemoteID} Specification}
\label{sec:remoteid}

The introduction of \acp{UAV}, their widespread adoption, and the increasing unsafe usage of these devices recently forced national and international aviation authorities to develop frameworks devoted to the regulation of their operations. In particular, authority bodies such as the US-based \ac{FAA} recently provided a set of standards and specifications which moderate and rule the traffic generated by civilians. This is the case of \emph{RemoteID}, a specification which provides means to the \ac{FAA} to integrate \acp{UAV} in the American aviation landscape~\cite{remoteID}. Specifically, \emph{RemoteID} provides the foundational regulations enabling the \ac{UAV} to communicate wirelessly with neighbouring entities, so as to enable control stations deployed by \acp{CI} to identify unsafe flying situations.

With \emph{RemoteID}, \acp{UAV} do not need to connect to the Internet to exchange information to a service supplier, but can create ad-hoc networks. In details, \acp{UAV} complying to the Remote ID specification must broadcast messages containing: a unique \ac{UAV} identifier, \ac{UAV} geographical information ( i.e., latitude, longitude, and altitude), \ac{UAV} speed, timestamp, and emergency status indicator.

\emph{RemoteID} envisions two technological solutions to allow \acp{UAV} to comply with the functional requirements. The first is via standard remote identification, where the \emph{RemoteID}-compliant \ac{UAV} transmits the above-listed information wirelessly using a radio module integrated into its design at the manufacturing time. The second is via a dedicated \emph{RemoteID} broadcast module, i.e., an additional module that can be attached to the \ac{UAV} at the deployment time to provide it with \emph{RemoteID} capabilities. In this latter case, the \ac{UAV} needs to add to the broadcast message the location of its take-off.
UAVs shall broadcast messages containing such information from take-off to shut down and, independently from the compliance method, with a minimum frequency of $1$~second using one of the available channels in the ISM frequency band (e.g., the $2.4$~GHz ISM band, using the WiFi communication technology).

Note that \emph{RemoteID} does not envision any cryptographic means to hide the information broadcasted in the wireless messages. Therefore, any user equipped with a compatible wireless receiver can detect and decode the messages, as well as infer information about the identity and location of the \ac{UAV}.

Finally, although today \emph{RemoteID} applies only for the American airspace, other regional regulatory bodies such as the EU~\cite{remoteid_europe} are also considering similar initiatives.

\subsection{\acl{DP} and \acl{Geo-Ind}}
\label{sec:diff_privacy}

The concept of \ac{DP} was initially formulated in the context of secure databases query, to protect information about the presence (or absence) of individual data. We hereby recall the definition of \ac{DP} in Definition~\ref{def:dp}, as provided by the authors in~\cite{dwork2014algorithmic}.

\begin{definition}[$\epsilon$-Differential Privacy] \label{def:dp}
Let $\mathcal{A}$ be a randomized algorithm that takes a dataset as input, and let $\text{im}(\mathcal{A})$ denote the image of $\mathcal{A}$. Given a positive real number $\epsilon$, the algorithm $\mathcal{A}$ is said to provide $\epsilon$-differential privacy if, for all dataset $D_1$ and $D_2$ differing on a single element, the following Eq.~\ref{eq:prob} holds for all $\mathcal{S} \subseteq \text{im}(\mathcal{A})$ and for a probability taken over the randomness of $\mathcal{A}$.
\begin{equation}
\label{eq:prob}
    Pr[\mathcal{A}(D_1) \in \mathcal{S}] \leq \exp(\epsilon) \, Pr[\mathcal{A}(D_2) \in \mathcal{S}]. 
\end{equation}
\end{definition}
The definition of \ac{DP} hence states that the probability that a query applied to two databases that differ from a single element returns a certain value is limited by $\exp(\epsilon)$. Thus, an adversary trying to infer the presence of an item in the two databases could formulate a guess having a minimum uncertainty bounded by  $\exp(\epsilon)$.

The concept of \ac{DP} has been further extended and exploited to preserve the privacy of location information \cite{machanavajjhala2008privacy, andres2013geo}. Let us consider a \ac{LBS} where a user may report location values from a set $\mathcal{P}$ of \ac{PoI}. Similarly to the process exploited for databases, researchers preserved the privacy of the location(s) of the user via a randomization algorithm $\mathcal{A}$ that perturbs location data by adding noise extracted from a statistical distribution. 

Instead of simply adding random noise, the authors in~\cite{andres2013geo} first proposed the concept of \acl{Geo-Ind}, as an extension to the notion of \ac{DP}. The rationale is that a user could have $(\epsilon, D)$-\ac{DP} within a radius $D$ from its location. The user specifies its privacy requirements via a tuple $(\ell,D)$, where $\ell$ represents the privacy requirements for radius $D$. Given $\epsilon=\ell/D$, the authors in~\cite{andres2013geo} provided the following definition~\ref{def:eGI}. 
\begin{definition}[$\epsilon$-\acl{Geo-Ind}]\label{def:eGI}
Let $x$ and $x'$ be two different location points, and let $d(x,x')$ be their Euclidean distance. Let $\mathcal{A}$ be a randomization algorithm.
Then, $\mathcal{A}$ satisfies $\epsilon$-\acl{Geo-Ind} if and only if the following Eq.~\ref{eq:dp} holds.
\begin{equation}
\label{eq:dp}
    Pr[\mathcal{A}(x) \in \mathcal{S}] \leq \exp \left (\frac{\ell}{D} d(x,x') \right )  \, Pr[\mathcal{A}(x') \in \mathcal{S}],  
\end{equation}
for all $\mathcal{S} \subseteq \text{im}(\mathcal{A})$, a probability taken over the randomness of $\mathcal{A}$, and for each couple $(x,x')$.
\end{definition}

We notice that Definition~\ref{def:eGI} is similar to Definition \ref{def:dp}, where the Euclidean distance replaces the Hamming distance between two databases. The replacement mentioned above is because Euclidean distance provides a relaxed privacy definition allowing to retrieve some information on the original data. 

In this paper, based on the results provided by~\cite{andres2013geo}, we consider the additive noise generated drawing samples from the \textit{planar Laplace distribution}, as per the following Definition~\ref{def:planLap}.
\begin{definition}[Planar Laplace Distribution]\label{def:planLap}
For a given actual location $x_{n,0} \in \mathrm{R}^2$ and a parameter $\epsilon \in \mathrm{R}^+$, the planar Laplace probability density function for each point $x \in \mathrm{R}^2$ is given by
\begin{equation}\label{eq:planLap}
    F_{\epsilon}^{x_{n,0}}(x) = \frac{\epsilon^2}{2 \pi} \exp\left( -\epsilon d(x_{n,0},x)\right) ,
\end{equation}
where $\epsilon^2/2 \pi$ is a normalization factor.
\end{definition}
The planar Laplace distribution has been analytically proven to provide \ac{Geo-Ind}~\cite{andres2013geo}. 

\section{Scenario and Adversary Model}
\label{sec:scenario_adv_model}

In this section, we first discuss the system model in \mbox{Sec.~\ref{sec:scenario}}. We then present the adversary model in Sec.~\ref{sec:adv_model}.

\subsection{Scenario and Assumptions}
\label{sec:scenario}

\begin{figure}[htbp!]
    \centering
    \includegraphics[width=.7\columnwidth]{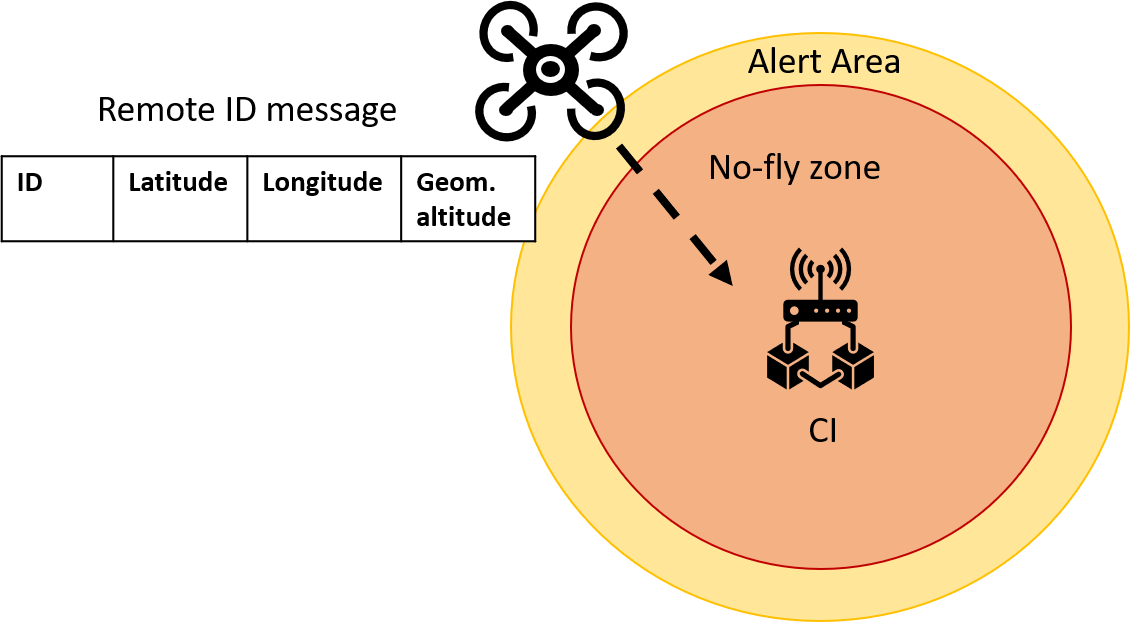}
    \caption{Overview of the considered scenario. 
    }
    \label{fig:scenario}
\end{figure}

Fig.~\ref{fig:scenario} depicts the scenario considered throughout this manuscript. In particular, we consider a scenario where a \ac{CI} needs to regulate the physical access to its proximity due to safety and privacy concerns. The \ac{CI} operator, namely $CI_i$, would like to identify invasions of the monitored area, namely the \emph{no-fly zone}, by unauthorized \acp{UAV}, to enhance people safety and prevent eavesdropping of sensitive information both in terms of audio and video recordings from suitably equipped \acp{UAV}. We hence define a safety radius $\delta_i$, originating from the centre location of the \ac{CI}. The safety radius identifies the region where \acp{UAV} represent a clear threat towards the integrity, privacy, and safety of the \ac{CI}, calling for the adoption of immediate actions. Although we consider a circular no-fly zone, our considerations apply to any shape of such area, including stretched shapes typical of airports runways.

Given that this manuscript aims to investigate the trade-off between privacy and detection capabilities provided by integrating \ac{Geo-Ind} techniques within the \emph{RemoteID} specification, our scenario assumes that the CI operator $CI_i$ deploys $J_i$ antennas to protect its site, \emph{all in the same location}, at the geographical centre of the CI deployment. This assumption is the worst case for the CI, formulated to analyze the behavior of the detection system in the worst condition, where the CI operator can rely only on \emph{RemoteID} messages emitted by the \acp{UAV} to identify invasions. Indeed, when $J_i \geq 1$, many antennas (possibly synchronized) are deployed in different locations, and thus, the detection system could adopt widespread time-based or RSSI-based wireless localization techniques to identify the location of the emitting UAV and improve its performance---this is out of the scope of this manuscript.

We assume that each antenna can receive a message from \acp{UAV} located in the circle centred at the antenna location and with radius $R_{\rm i,max}$. The value $R_{\rm i,max}$ depends on the physical environment and the physical features of the transmitted signal (e.g., the carrier frequency), and it is also dynamic. Without loss of generality, we assume that \acp{UAV} emit the \emph{RemoteID} messages using the WiFi communication technology on a specific channel. Thus, we assume that the antennas are tuned to eavesdrop on messages transmitted over WiFi. This latter is also a realistic assumption, as the \emph{RemoteID} specification indicates that the used channel(s) should be known to the general public.

In line with Fig.~\ref{fig:scenario}, we also assume that $\delta_i < R_{\rm i,max}$, and that the circular crown centered in the location of the CI monitoring station(s), with radius $r_i = \delta_i - R_{\rm i,max}$, is defined as the \emph{alert area}. \acp{UAV} detected when located in the \emph{alert area} do not lead to immediate action by the CI operator, as flying within this area is allowed. However, being the \emph{alert area} close to the \emph{no-fly zone}, the CI operator goes in an alert state, ready to take actions as soon as an invasion of the \emph{no-fly zone} is detected. Overall, the \emph{alert area} models the unpredictability of the receiving range of the wireless antenna, and a careful design of the monitoring system requires the \emph{alert area} to be very small. Indeed, the CI operator is mainly interested in monitoring invasion of the \emph{no-fly zone}. 
Thus, the CI operator would like to have $r_i = 0$, i.e., $\delta_i = R_{\rm i,max}$, which is impossible to achieve in practice due to the unpredictability and dynamicity of the receiving range.
Overall, the CI operator aims at detecting, through the information contained in the RemoteID messages and acquired via its monitoring stations, whether any \acp{UAV} invade the \emph{no-fly zone} to take necessary actions. 

On the \acp{UAV} side, as will be clarified in Sec.~\ref{sec:DiPrID}, we assume that \acp{UAV} exploit \ac{DP} mechanisms to hide their actual location $x_{n,0}$ from non-legitimate receivers. Thus, the \emph{RemoteID} messages broadcasted by the \acp{UAV} include obfuscated locations, hiding the exact UAV's location. Moreover, we assume that \acp{UAV} also emit pseudo-anonymous messages, where the true identity is obfuscated to the general public and can be recovered only by a TTP, such as the FAA. This service can be achieved thanks to the usage of \emph{pseudonyms}, that can be either generated on the fly~\cite{tedeschi2021_acsac} or provided by a TTP~\cite{rajput2016_access}. The specific way such pseudonyms are issued and managed is out of the scope of this manuscript. We also assume that the \acp{UAV} have the computational power necessary to run asymmetric encryption and decryption algorithms and to store a 128-bit symmetric key, in line with the capabilities of modern \acp{UAV}.

We also assume the presence of a \ac{TTP}, reachable by the operators of the \acp{UAV} and the \acp{CI} via the Internet. The role of the TTP is first to register all \acp{UAV} and \ac{CI} operators (see Sec.~\ref{sec:registration}). Then, as described in Sec.~\ref{sec:reporting}, in case of invasion detected by a CI operator, the TTP can verify such invasion and provide information about the identity and current location of the invading UAV. In the real world, the role of the TTP can be played either by the US-based FAA, or by additional regional entities, or by further institutions, delegated by the regional ones.
Finally, we report the main notation used in this paper in the Annex.

\subsection{Adversary Model}
\label{sec:adv_model}

In this paper, we consider different attacker models, characterized by both passive and active capabilities, as described below.

{\bf Unaware Invader, $\mathcal{A}_1$.} This is in principle an honest user, who is flying the drone in the nearby of the \ac{CI} and, without intention, invades the no-fly zone, behaving as an attacker. This user is usually unskilled: thus, he/she is not able to tamper with the \ac{UAV} firmware, e.g., $\mathcal{A}_1$ cannot send desired bogus location information. The latter is a reasonable assumption, as most of the area invasions attacks detected by CI operators in the last years were performed by unskilled users, simply flying their \acp{UAV} too close to the CI deployment~\cite{dedrone}.

{\bf Rookie Pilot Attacker, $\mathcal{A}_2$. } This attacker flies intentionally a drone towards the \ac{CI}, either to gather sensitive information or to cause physical damage to the CI site (e.g., drop a bomb). Similarly to the previous model, we assume that $\mathcal{A}_2$ is not able to tamper with the \ac{UAV} firmware, e.g., $\mathcal{A}_2$ cannot send desired bogus location information. This latter is also a reasonable assumption, as many criminal organizations simply use off-the-shelf drones for their malicious purposes, without having the capabilities to tamper with the drone.
\noindent
We assume that the attacker $\mathcal{A}_2$ may want to jeopardize the confidentiality and the availability of the services provided by the \ac{CI}. In this case, the \ac{UAV} is equipped with cameras to record images \cite{zhi2020security}, microphones to record acoustic emissions, or with modules to cause disruption on purpose, such as jamming \cite{bhattacharya2010differential}, to threaten the \ac{CI} availability.

{\bf Eavesdropper, $\mathcal{A}_3$.} We assume that the objective of $\mathcal{A}_3$ is to track a certain victim \ac{UAV}. Note that attackers can exploit this process for multiple purposes. Due to the unmanned nature of \acp{UAV}, an attacker able to track a delivery \ac{UAV} may physically capture it and steal the carried load. This attack possibly causes both physical damage to the \ac{UAV} and a monetary loss, due to the theft of the package. Furthermore, the leakage of the location of the \ac{UAV} may represent a threat also due to those that are ``enraged by drones".  Lastly, a company may want to track \acp{UAV} belonging to a competitor company to cause financial damages. We assume that $\mathcal{A}_3$ has enough memory and computational power to collect and store successive \emph{RemoteID} messages and to derive statistics from them. Based on the location information, the attacker may be able to track a specific \ac{UAV} thanks to the presence of the unique \ac{UAV} identifier contained in each \emph{RemoteID} message. Furthermore, the attacker may be able to predict the trajectory of the target \ac{UAV} to cause further damages. Note that, to avoid detecting its presence based on its identity information, the \ac{UAV} may send messages with a pseudonym, as previously highlighted~~\cite{babaghayou2020pseudonym}. However, this might not be sufficient to protect against tracking, as the attacker may be able to infer the identity of the \ac{UAV} based on the plain-text location information \cite{emara2013vehicle,zhang2020novel}. Furthermore, notice that the unique identifier included in \emph{RemoteID} messages indicates that the connection between the \ac{UAV} ID and the physical owner will never be disclosed. However, we argue that an attacker may visually recognize a target \ac{UAV} (e.g., delivery \ac{UAV} of a specific company) and simultaneously capture the \emph{RemoteID} traffic to infer its ID. Also, note that the adoption of rolling acronyms allows de-correlating packets emitted by the drone. Thus, even deploying a network of distributed synchronized sensors on the ground, the attacker cannot reliably link multiple packets to the same emitter, drastically reducing the effectiveness of multi-lateration localization techniques.

It is worth noting that if the attacker has the capabilities to switch off RemoteID or tamper with the firmware of the UAV (thus being not a rookie), our study is not applicable. At the same time, we remark that this manuscript aims at investigating the trade-off between location privacy of UAVs and data utility, and not at detecting any kind of invading drone. Thus, such an advanced adversary model is out of the scope of our analysis, and can be addressed through already-available techniques (e.g, audio or radar).

In a nutshell, the techniques described in this manuscript intend to allow the \acp{UAV} to obtain an enhanced level of location privacy at the cost of a few false invasions detected by the CI operator. The TTP will assist the CI in determining if the invasion actually occurred and, if verified, it will provide the actual location of the invading device. We will give more details below.

\section{Preserving Location Privacy and Utility in \acp{UAV} Detection}
\label{sec:proposals}

In this section, we first illustrate the \emph{Registration Phase} (Sec.~\ref{sec:registration}). Then, Sec.~\ref{sec:DiPrID} explains DiPrID, Sec.~\ref{sec:loc_reps} introduces encrypted location reports, Sec.~\ref{sec:detection} describes the proposed area invasion detection technique and, finally, Sec.~\ref{sec:reporting} details the \emph{Reporting Phase}.

\subsection{Registration Phase}
\label{sec:registration}

In the \emph{Registration Phase}, executed before deployment, both the CI operators and the UAV register with the TTP, e.g., the FAA. For this phase, we assume that the TTP exposes a dedicated web interface, reachable by both UAV manufacturers and CI operators via Internet and secured via well-known protocols (e.g., SSL/TLS).

\begin{figure}
    \centering
    \resizebox{.8\columnwidth}{!}{%
    
    \begin{tikzpicture}[box/.style={draw,fill=white,align=center}, scale=0.9, ]

    \draw [line width=1mm] (0,0) -- ++(1,0) coordinate[midway](A)
    (7,0) -- ++(1,0) coordinate[midway](B);

    \draw (A) -- ++ (0,8) 
    node[fill=white,draw=black,minimum width=2cm,rotate=0, pos=1, label={[label distance=0.6cm, scale=1.3]:Operator UAV $n$}] (AH) {}
    
    node[box,pos=0.75, fill=blue!5, align=left, scale=1.3] (A1) {Generate $ID_n$.}
    
    node[box,pos=0.2, fill=blue!5, align=left, scale=1.3] (A2) {Store $P_{TTP}$}
    
    coordinate[pos=0.88] (XX1) coordinate[pos=0.88] (XX2) coordinate[pos=0.65] (X1) coordinate[pos=0.65] (X2);
    
    \draw[latex'-latex', double, thick] (XX1 -| A) -- (XX2 -| B) node[midway, above, scale=1.3]{Secure Connection Setup};
    
    \draw[-latex,dashed,thick] (X1) -- (X2 -| B) node[midway,above,scale=1.3]{$ID_n$};

    \draw (B) -- ++ (0,8) 
    node[fill=white,draw=black,minimum width=2cm,rotate=0, pos=1, label={[label distance=0.6cm,scale=1.3]:TTP}] (BH) {}
    
    node[box,pos=0.5, fill=blue!5, align=left,scale=1.3] (B1) {Create entry in UAV Table: $< ID_n>$ }
    
    coordinate[pos=0.3] (X1) coordinate[pos=0.3] (X2);
    
    \draw[-latex,dashed,thick] (X2) -- (X1 -| A) node[midway,above,,scale=1.3]{$P_{TTP}$};

    \end{tikzpicture}
    }
    \caption{\emph{Registration Phase} for the \acp{UAV}.}
    \label{fig:reg_phase_uav}
\end{figure}
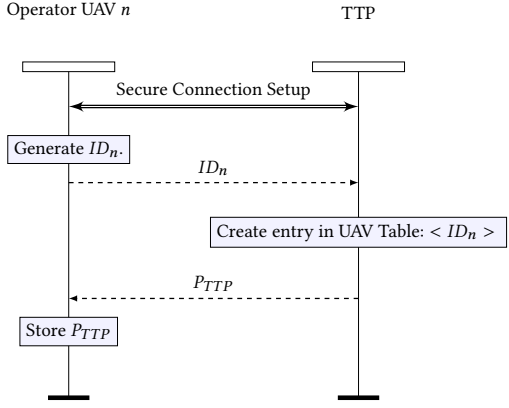

The information delivered to the TTP is different based on the nature of the communicating party, being it a CI operator or a UAV manufacturer. As depicted in Fig.~\ref{fig:reg_phase_uav}, if the entity registering with the TTP is a UAV, after the secure connection setup, the manufacturer submits the unique ID $ID_n$ of the UAV, to be used also for generating the acronyms used into \emph{RemoteID} messages. The TTP stores the unique ID of the UAV in a local table, namely the \emph{\acp{UAV} Table}, and it provides in response to the UAV its public key $P_{TTP}$, used by the UAV when generating encrypted location reports. 

\begin{figure}
    \centering
    \resizebox{.8\columnwidth}{!}{%
    
    \begin{tikzpicture}[box/.style={draw,fill=white,align=center}, scale=0.9, ]

    \draw [line width=1mm] (0,0) -- ++(1,0) coordinate[midway](A)
    (7,0) -- ++(1,0) coordinate[midway](B);

    \draw (A) -- ++ (0,8) 
    node[fill=white,draw=black,minimum width=2cm,rotate=0, pos=1, label={[label distance=0.6cm, scale=1.3]:CI Operator $i$}] (AH) {}
    
    node[box,pos=0.75, fill=blue!5, align=left, scale=1.3] (A1) {Generate Location $\mathbf{pos_i}$\\Generate Radius no-fly zone $\delta_i$.}
    
    node[box,pos=0.15, fill=blue!5, align=left, scale=1.3] (A2) {Store $CI_i$}
    
    coordinate[pos=0.88] (XX1) coordinate[pos=0.88] (XX2) coordinate[pos=0.56] (X1) coordinate[pos=0.56] (X2);
    
    \draw[latex'-latex', double, thick] (XX1 -| A) -- (XX2 -| B) node[midway, above, scale=1.3]{Secure Connection Setup};
    
    \draw[-latex,dashed,thick] (X1) -- (X2 -| B) node[midway,above,scale=1.3]{$\mathbf{pos_i}$, $\delta_i$};

    \draw (B) -- ++ (0,8) 
    node[fill=white,draw=black,minimum width=2cm,rotate=0, pos=1, label={[label distance=0.6cm,scale=1.3]:TTP}] (BH) {}
    
    node[box,pos=0.45, fill=blue!5, align=left,scale=1.3] (B1) {Create entry in CI Table:\\ $< CI_i, \mathbf{pos_i}, \delta_i>$ }
    
    coordinate[pos=0.27] (X1) coordinate[pos=0.27] (X2);
    
    \draw[-latex,dashed,thick] (X2) -- (X1 -| A) node[midway,above,,scale=1.3]{$CI_i$};

    \end{tikzpicture}
    }
    \caption{\emph{Registration Phase} for the \ac{CI} operator.}
    \label{fig:reg_phase_ci}
\end{figure}
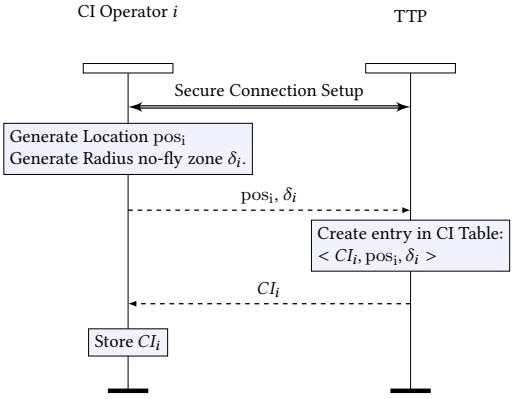

If the entity registering with the TTP is a CI operator, as shown in Fig.~\ref{fig:reg_phase_ci}, the CI operator submits: (i) the location of the CI to be protected (expressed in latitude, longitude, and altitude), namely $\mathbf{pos_i}$, and (ii) the radius of the no-fly zone (in meters), namely $\delta_i$. The FAA stores such information in a dedicated table, namely the \emph{CI Table}, together with a unique identifier $CI_i$. Such identifier is provided in response to the CI operator, to be used later on in case of invasion, during the \emph{Reporting Phase} (see Sec.~\ref{sec:reporting}).

Note that the \emph{Registration Phase} should be executed only once per each UAV and CI site, before their deployment. The CI operator could decide to update any of the submitted information later on, and this could be possible thanks to the usage of the unique CI identifier $CI_i$. Also, the UAV could decide to update their information, e.g., in case of a change of the identifier $ID_n$. Similar to the CI operators, \acp{UAV} could require an update of the information by submitting their previous identifiers $ID_n$ and authenticating with the TTP. The specific way the authentication is performed via the Internet is out of the scope of this manuscript.

\subsection{Differentially-Private RemoteID}
\label{sec:DiPrID}

In this section, we introduce our solution to enhance location privacy on \acp{UAV} compliant with the \emph{RemoteID} specification, namely \emph{Differentially-Private RemoteID} (DiPrID), based on the notion of \ac{Geo-Ind}. To this aim, we exploit the definition of the planar Laplace distribution reported in Def.~\ref{def:planLap}. 

The main intuition used to draw a random point from the planar Laplace distribution is that such distribution only depends on the distance from the random variable realization and the real \ac{UAV} location $x_{n,0}$. 
Therefore, we obtain a convenient representation of the planar Laplace distribution by exploiting the polar coordinates system, in line with the proposal by the authors in~\cite{andres2013geo}. Hence, Eq.~\eqref{eq:planLap} can be rewritten as in Eq.~\ref{eq:planLap_2}.
\begin{equation}
\label{eq:planLap_2}
    F_{\epsilon}(D,\theta) = \frac{\epsilon^2}{2 \pi} D\exp\left( -\epsilon D \right), 
\end{equation}
where we considered the Laplace distribution centered at $x_{n,0}$, $D$ is the distance between a generic value $x$ and $x_{n,0}$, and $\theta$ is the angle that the segment joining $x$ and $x_{n,0}$ forms with the horizontal axis.

The variables $D$ and $\theta$ are statistically independent~\cite{andres2013geo}. Therefore, we obtain $\theta$ by drawing a random point from a uniform distribution between $0$ and $2 \pi$, and $D$ via the following Eq.~\ref{eq:D}.~\cite{andres2013geo}
\begin{equation}\label{eq:D}
    D = -\frac{1}{\epsilon} \left ( \gamma_{-1}\left( \frac{p-1}{e} \right)+ 1 \right),
\end{equation}
where $p$ is drawn from a uniform distribution in $[0,1)$ and $\gamma_{-1}$ is the $-1$ branch of the Lambert function \cite{olver2010nist}. 

Let us denote as $x_{n,0}^m$ the vector containing the \ac{UAV} location information, i.e., its latitude and longitude at time instant $m$ (the altitude, having different range, has to be considered on its own). 
Then, the \ac{UAV} applies \ac{DP} such that any receiver receives the perturbed \ac{UAV} state information in Eq.~\ref{eq:perturbed}.
\begin{equation}
\label{eq:perturbed}
    x_n^m = x_{n,0}^m + <D^m \cos(\theta^m), D^m \sin(\theta^m)>,
\end{equation}
where $D^m$ and $\theta^m$  are the generated parameters of the planar Laplace distribution, while $m$ indicates the time instant.

Since this mechanism potentially generates points everywhere in the plane, we apply truncation to confine the perturbed locations in a certain area~\cite{andres2013geo}. We hence define the set $\mathcal{L}$ of admissible locations independent from the actual location $x_{n,0}$. Therefore, we apply Eq.~\ref{eq:bound}
\begin{equation}
\label{eq:bound}
    x_n^m = 
    \begin{cases}
    x_n^m, \, \text{if} \, x_n^m \in\mathcal{L}; \\
    \min\left ({x_n^m}, \mathcal{L}\right ), \, \text{if} \, x_n^m \notin \mathcal{L};
    \end{cases}
\end{equation}
where $\min\left (x_n^m, \mathcal{L}\right )$ indicates the point in $\mathcal{L}$ closest to $x_n^m$. The truncation procedure satisfies $\epsilon$-\ac{Geo-Ind}~\cite{andres2013geo}.

Algorithm \ref{alg:geoInd} shows the steps needed to compute the obfuscated location value via the proposed DiPrID algorithm.
\begin{algorithm}
\footnotesize
{\bf let } $x_{n,0}^m$ be the actual location at time $m$\;
{\bf let} $\epsilon$, $D$ be the parameters of the Laplace distribution\;
{\bf let} $\mathcal{L}$ be the set of admissible locations\;
\BlankLine
\tcc{Draw a random value from the planar Laplace distribution}
draw $\theta$ from a uniform distribution $\in [0,2\pi]$\;
draw $p$ from a uniform distribution $\in [0,1]$\;
generate $D^m$ from \eqref{eq:D}\;
\tcc{Obfuscate value}
compute $x_n^m$ via \eqref{eq:perturbed}\;
\tcc{Apply truncation}
compute truncated $x_n^m$ via \eqref{eq:bound}\;
\KwResult{$x_n^m$\;}
\caption{Pseudo-code of the obfuscation technique.}
\label{alg:geoInd}
\end{algorithm}

Note that the procedure detailed above applies also for the altitude at the time instant $m$. However, such a parameter is handled individually, with its own values of $\epsilon$ and $D$, due to the different range it has compared to latitude and longitude.

We notice that the mechanism mentioned above truly guarantees $\epsilon$-\ac{Geo-Ind} only when considering the disclosure of independent locations. However, this is not the case of \emph{RemoteID}, where locations are disclosed with a maximum periodicity of $1$~second, hence being correlated. In this case, the effective degree of location privacy of the UAV depends on the number of disclosed locations, and specifically, on the number of messages received by the CI operator, that is unknown to the emitter. Without loss of generality, assuming $k$ messages are received by a given CI monitoring station, the overall degree of location privacy of the UAV is $k \cdot \epsilon$, as demonstrated by the authors in~\cite{andres2013geo}. Enhanced location privacy could be provided by integrating variations of the \ac{Geo-Ind} scheme, such as the one proposed by the authors in~\cite{xiao2015protecting}, where temporal correlations are considered and compensated for. We recall that our contribution aims to show that enhanced location privacy can be provided to \acp{UAV} compliant with the \emph{RemoteID} rule, without nullifying detection efforts. On the one hand, we believe that providing $k \epsilon$ privacy significantly increases the location privacy, compared to the actual situation, where no location privacy at all is considered. On the other hand, as part of our future work, we will consider solutions able to increase location privacy beyond $k \epsilon$.

\subsection{Encrypted Location Reports}
\label{sec:loc_reps}

Although enhancing location privacy for \acp{UAV}, DiPrID limits the capabilities of the CI operators to localize possibly invading \acp{UAV}.
To this aim, we include in \emph{RemoteID} messages \emph{Encrypted Location Reports}, i.e., ciphertexts encoding precise location information.

From Sec.~\ref{sec:registration}, recall that the generic UAV $d_n$ stores the public key of the TTP, namely $P_{TTP}$.
Every time the UAV needs to deliver a new \emph{RemoteID} message, besides applying the DiPrID technique described in Sec.~\ref{sec:DiPrID}, it also generates a one-time key $K_{n}$. Using such a key, $d_n$ encrypts its actual identity $ID_n$ and location $x_{n,0}^m$, as:
\begin{equation}
    \label{eq:report}
    c_{n,0}^m = \tilde{E} \left( ID_n || x_{n,0}^m, K_{n}  \right),
\end{equation}
where $\tilde{E} \left( \cdot \right)$ refers to a generic symmetric encryption algorithm (such as AES, widely supported on commercial \acp{UAV}). Then, using the public key of the TTP, it generates the encrypted key $\tilde{K}_{n}$, as
\begin{equation}
    \label{eq:encr_key}
    \tilde{K}_{n} = \tilde{E}_S \left( K_{n}, P_{TTP} \right),
\end{equation}
where $\tilde{E}_S$ is a generic asymmetric encryption algorithm, such as RSA. The resulting values, i.e., $\tilde{K}_{n}$ and $c_{n,0}^m$, form the \emph{encrypted location report} delivered within the corresponding \emph{RemoteID} message.

As will be described in Sec.~\ref{sec:reporting}, such location reports will be useful in case of invasion detection, to let the TTP extract the actual location and identity of the invading \acp{UAV}.

\subsection{Area Invasion Detection}
\label{sec:detection}

In this section, we describe ICARUS, i.e., the area invasion detection strategy implemented by a generic \ac{CI} monitoring station to detect invasions of the no-fly zone by unauthorized \acp{UAV} compliant with the \emph{DiPrID} extension described in Sec.~\ref{sec:DiPrID}.

We report in~Algo.~\ref{alg:detection} the pseudo-code of the ICARUS algorithm, while each step is described and motivated below.

\begin{algorithm}
\footnotesize
{\bf let } $R_{\rm i,max}$ be the Reception range of $CI_i$\;
{\bf let} $\delta_i \leftarrow [0, R_{\rm i,max}]$ be the Detection Range\;
{\bf let} $geo\_pos_i = [lat_i,lon_i,alt_i ]$ be the location of the receiver in geographical coordinates\;
{\bf let} $xyz\_pos_i = [x_i, y_i, z_i$] be the equivalent location of the receiver $i$ in Cartesian coordinates\;
{\bf let } $W \leftarrow [0, W_{MAX}]$ be the Detection Window\;
{\bf let} $W_{en} = \{ 0,1\}$ a flag indicating if the evaluation of $W$ already started\;
\BlankLine
$W_{en} = 0$\;
\Do{( A \emph{RemoteID} packet $msg_m$ is received) || ($W$ elapsed).}{Listen on the wireless channel (WiFi)\;}
\BlankLine
\If{$W$ elapsed}{\KwResult{Invasion=0\;}}
\Else{
\tcc{A packet $msg_m$ was received}
$W_{en}=1$\;
\tcc{Extract relevant info from $msg_m$}
$msg_m$ $\rightarrow$ {$ID_m$, $t_m$, $lat_m$, $lon_m$, $alt_m$, $v\_x_{m}$, $v\_y_{m}$, $v\_z_{m}$ }\;
$geo\_pos_m$ = $[lat_m,lon_m,alt_m ]$\;
$xyz\_pos_m$ = $[x_m, y_m, z_m]$\;
\tcc{Compute Distance}
$d_{i,m} = norm(xyz\_pos_i - xyz\_pos_m)$\;
\BlankLine
\tcc{Apply Detection Strategy}
    \If{$d_{i,m} < \delta_i$}{
        \tcc{Invasion Detected}
        $det\_events[m] = 1$\;
        \KwResult{Invasion=1}
    } \Else{
        \tcc{Invasion NOT Detected}
        $det\_events[m] = 0$\;
        \If{$t_m - t_0 \ge W$}{
        \KwResult{Invasion=0}
        } \Else{
        Wait until $W$\;
        }
    }
}
\caption{ICARUS area invasion detection algorithm.}
\label{alg:detection}
\end{algorithm}

We now introduce the notion of the \emph{Detection Window}, $W$.
\begin{definition}[Detection Window $W$] \label{def:W} 
The \emph{Detection Window}, referred to as $W$, is defined as the maximum time interval where the detection system needs to provide indication of an \emph{invasion} (or \emph{not invasion}) of the no-fly zone by any \acp{UAV}.
\end{definition}
Note that the detection window is used as a time reference, to indicate the presence of an invasion in a given time.

According to ICARUS, a receiver deployed to monitor the sensitive area continuously listens on the wireless channel, waiting for incoming \emph{RemoteID} messages (lines 8--10). At most every $W$~seconds, ICARUS provides a decision about the presence of an invading \ac{UAV} based on the evaluation of the received \emph{RemoteID} messages (lines 11--13). Let us define $t_0$ as the time the detection window starts.
As soon as a \emph{RemoteID} message is received, the CI monitoring station extracts the parameters necessary for area invasion evaluation. They include: (i) the (pseudo)-ID of the sender, $ID_m$; (ii) the timestamp of the message $t_m$; (iii) the geographical coordinates of the emitter, $lat_m$, $lon_m$, $alt_m$, and finally (iv) the speed components $v\_{x_m}$, $v\_{y_m}$, and $v\_{z_m}$ (line 15).

Let us define as $xyz\_pos_i$ the equivalent location of the $i$-th receiver in a 3-D Cartesian coordinates plane obtained from the traditional geographical coordinates (latitude, longitude, altitude). Based on the location reported in the message, the CI monitoring station can compute the distance between the monitoring station and the emitter, namely $d_{i,m}$ (line 18). 

Then, the CI monitoring station applies the following detection strategy (lines 19--30). If the reported location is within the detection threshold $\delta_i$, an invasion is detected, and ICARUS immediately returns the \emph{invasion} result (lines 19--21). Note that this is the most conservative behaviour that the CI monitoring station can keep, i.e., a single invasion event leads to the conclusion that an invasion is ongoing. Although more complex definitions of \emph{invasion} could be thought, they could potentially lead to reduced false positives, at the expense of chances of reduced true positives, which is not desirable for a CI monitoring station.
Otherwise, if the distance is $ > \delta_i$ but $<R_{\rm i,max}$, if the timestamp reported in the message is such that $W$ still has to expire (lines 24--25), the location is stored for future use. If $W$ expires, ICARUS provides the result \emph{not invasion}, indicating that no invasion has been detected during the last $W$~seconds.

To sum up, as soon as a single \emph{invasion} event is detected, ICARUS returns an \emph{invasion} output. Conversely, to return \emph{not invasion}, all the messages received within $W$ must report a \emph{not invasion} event.

If ICARUS returns \emph{invasion} as output, the following \emph{Reporting Phase} is triggered by the CI operator. Otherwise, no action is taken.

\subsection{Reporting Phase}
\label{sec:reporting}

In the \emph{Reporting Phase}, the CI operator reports the UAV invasion(s) to the TTP, asking for the actual location of the UAV, so to trigger the necessary actions. Note that the TTP will provide the actual location of the invading entity only if the UAV is actually invading the no-fly zone of the CI. Otherwise, no location nor identity is provided back to the CI, still preserving UAV privacy.

\begin{figure}
    \centering
    \resizebox{.8\columnwidth}{!}{%
    
    \begin{tikzpicture}[box/.style={draw,fill=white,align=center}, scale=0.9, ]

    \draw [line width=1mm] (0,0) -- ++(1,0) coordinate[midway](A)
    (7,0) -- ++(1,0) coordinate[midway](B);

    \draw (A) -- ++ (0,8) 
    node[fill=white,draw=black,minimum width=2cm,rotate=0, pos=1, label={[label distance=0.6cm, scale=1.3]:CI Operator $i$}] (AH) {}
    
    node[box,pos=0.75, fill=blue!5, align=left, scale=1.3] (A1) {Invasion Detected}
    
    node[box,pos=0.1, fill=blue!5, align=left, scale=1.3] (A2) {Take countermeasures}
    
    coordinate[pos=0.88] (XX1) coordinate[pos=0.88] (XX2) coordinate[pos=0.56] (X1) coordinate[pos=0.56] (X2);
    
    \draw[latex'-latex', double, thick] (XX1 -| A) -- (XX2 -| B) node[midway, above, scale=1.3]{Secure Connection Setup};
    
    \draw[-latex,dashed,thick] (X1) -- (X2 -| B) node[midway,above,scale=1.3]{$CI_i$, $\left( ID_{n}, geo\_pos_m, \dots, c_{n,0}^m \right)$};

    \draw (B) -- ++ (0,8) 
    node[fill=white,draw=black,minimum width=2cm,rotate=0, pos=1, label={[label distance=0.6cm,scale=1.3]:TTP}] (BH) {}
    
    node[box,pos=0.4, fill=blue!5, align=left,scale=1.3] (B1) {Retrieve $CI_i$\\Verify $|| pos_i - pos_m || < \delta_i$\\ $K_{n}$ = $\tilde{D}_S \left( \tilde{K}_{n} , p_{TTP} \right)$ \\ $x_{n,0}^m = \tilde{D} \left( c_{n,0}^m, K_{n}  \right)$ }
    
    coordinate[pos=0.17] (X1) coordinate[pos=0.17] (X2);
    
    \draw[-latex,dashed,thick] (X2) -- (X1 -| A) node[midway,above,,scale=1.3]{$x_{n,0}^m$};

    \end{tikzpicture}
    }
    \caption{Sequence diagram of the \emph{Reporting Phase}.}
    \label{fig:reporting}
\end{figure}
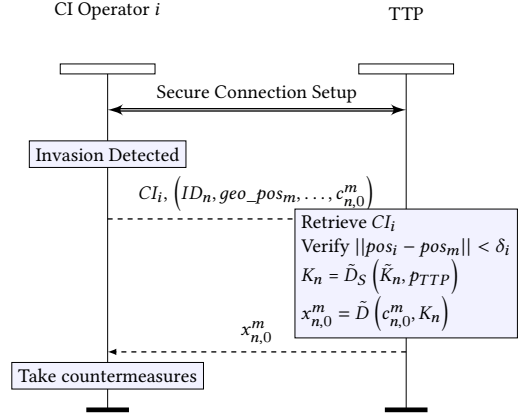

The sequence diagram of the operations executed by the CI operator and the TTP during the \emph{Reporting Phase} is shown in Fig.~\ref{fig:reporting}.

Specifically, this phase assumes that: (i) the CI operator has previously registered with the TTP, providing its identifier, location, and details of the \emph{no-fly} zone; (ii) the TTP exposes a web interface, secured via the well-known SSL/TLS protocol, where CI operators could submit reports about UAV invasions; and (iii) the connection between the TTP and the CI operators is a low-delay connection.

As soon as an invasion is detected, the CI operator reports to the TTP the following information: (i) the ID of the CI operator, obtained from the registration; and (ii) the \emph{RemoteID} message(s) emitted by the UAV and causing the invasion of the \emph{no-fly zone}.
The TTP first verifies the occurrence of the invasion based on the publicly available informed contained in the \emph{RemoteID} messages, i.e., that the position reported in clear-text by the UAV (obfuscated through the DiPrID technique) causes an invasion of the no-fly zone of the CI operator. If the \emph{obfuscated} location reported by the UAV is verified to lead to an invasion, the TTP decrypts the \emph{encrypted location reports} of the \emph{RemoteID} messages. Thus, the TTP first extracts the one-time key $K_n$, by decrypting the value $\tilde{K}_{n}$, as: 
\begin{equation}
    \label{eq:decr_key}
    K_{n} = \tilde{D}_S \left( \tilde{K}_{n} , p_{TTP} \right),
\end{equation}
where $\tilde{D}_S$ is the asymmetric decryption algorithm dual of $\tilde{E}_S$ (e.g., RSA), using the private key of the TTP $p_{TTP}$. Then, using $K_{n}$, the TTP decrypts the location information included in the message as 
\begin{equation}
    \label{eq:decrypt}
     x_{n,0}^m = \tilde{D} \left( c_{n,0}^m, K_{n}  \right),
\end{equation}
where $\tilde{D} \left( \cdot \right)$ refers to a symmetric decryption algorithm (if AES was used for encryption, $D$ will be the AES decryption algorithm). Eqs.~\ref{eq:decr_key} and ~\ref{eq:decrypt} are applied for every message reporting the invasion of the \emph{no-fly zone} of the CI operator.

Then, the TTP can verify if the actual location of the UAV is inside the no-fly zone of the CI. Only if such a condition is verified, the decrypted location(s) are finally delivered to the CI operator, allowing it to know the exact location of the invading UAV and take action. 
Conversely, if the invasion is not verified, no location is provided back by the TTP to the CI operator, so preserving the location privacy of the UAV. Similarly, no actions will be taken by the TTP if the CI operator is not registered and known to the TTP. 

\section{Performance Evaluation}
\label{sec:performance}

We present the dataset and the simulation setup in Sec.~\ref{sec:dataset}, while we report and discuss the results of our analysis in Sec.~\ref{sec:simulations}.

\subsection{Reference Dataset and Experiments Setup}
\label{sec:dataset}

To analyze the implications of the proposed detection strategy and the related privacy properties offered to \acp{UAV}, we run an extensive simulation campaign using the tool Matlab R2020b.

\begin{figure}[htbp!]
    \centering
    \includegraphics[width=0.7\columnwidth]{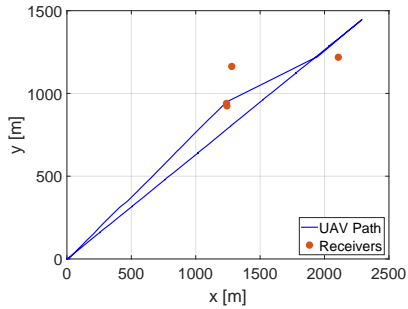}
    \caption{2-D representation of our considered scenario.}
    \label{fig:experiments}
\end{figure}

For the simulations, we considered data related to the flight of \acp{UAV} and the location of the receivers released freely and open-source by the NATO's Emerging Security Challenges Division, available at~\cite{dataset}. The data have been originally provided in the context of a challenge, where the participants' task was to track, classify, and identify Class I UAVs as they fly within a defined area from the data provided by the available sensors. The data include the log files of the UAV providing, among the others, information about the specific location (latitude, longitude, and altitude) of the UAV at a given time (reported through a timestamp with a precision of $1$~$\mu$s), and the instantaneous readings of the speed of the UAV (along the three-axis $x$-$y$-$z$). Such information is available with an average frequency of $90$~msec, well below the maximum reporting time of $1$~second imposed by the \emph{RemoteID} specification. As the values mentioned above are the same to be reported through \emph{RemoteID} messages, we assumed that the above values are delivered within wireless messages broadcasted on the wireless channel. The dataset also includes the location of four ($4$) receivers deployed randomly in the scenario. Overall, the area travelled by the drone in the experiments is $1.5 \times 2.5 $~km, and it is shown in Fig.~\ref{fig:experiments}. We also considered a maximum reception range $R_{\bm i,max} = 705$~meters on the CI monitoring station and the safety radius $\delta_i$ of the no-fly zone equal to $\delta_i=700$~meters. Thus, the radius $r_i$ of the \emph{alert area} is $r_i=R_{\bm i,max}-\delta_i=5$~meters.

Within the simulations, we converted all the geographical coordinates of the above-described entities to an equivalent Cartesian coordinate system. Then, for each timestamp where information on the UAV is available, we computed the distance between the current location of the UAV and the receiver's location. In line with Algo.~\ref{alg:detection}, if such a distance is less than $R_{\bm i,max}$, we considered the related information to be available to the specific receiver (to be used for invasion detection purposes). Otherwise, due to the physical limitations of the adopted communication technology, we assumed the \emph{RemoteID} messages as lost by the specific receiver, and thus, the related information as \emph{not} available. As per Algo.~\ref{alg:detection}, each receiver is independent from other receivers in the area.

In line with the methodology described in Sec.~\ref{sec:DiPrID}, we perturbed the true location of the UAV with values extracted at random from a planar Laplacian distribution, and we assumed the perturbed values to be the ones delivered to the receivers. On the detection side, using perturbed information, we applied the detection technique described in Sec.~\ref{sec:detection} and evaluate the capability of the detection system to correctly and timely identify area invasion.
Finally, note that all our results have been obtained through simulations run on an HP ZBook Studio G5, equipped with two Intel Core i7-9750H processors running at $2.60$~GHz, $32$~GB of RAM, and $1$~TB of HDD.

\subsection{Experimental Analysis}
\label{sec:simulations}

All the results presented here have been reported as average values over $10,000$~runs, where for each run, we extracted different random values of the Laplacian distribution. As all the receivers provided very similar results, we independently averaged the results over all the $4$ receivers contained in the dataset, considering the whole \ac{UAV} trajectory depicted in Fig.~\ref{fig:experiments}, lasting about $13$~minutes.

\begin{figure}[htbp!]
    \centering
    \includegraphics[width=.8\columnwidth]{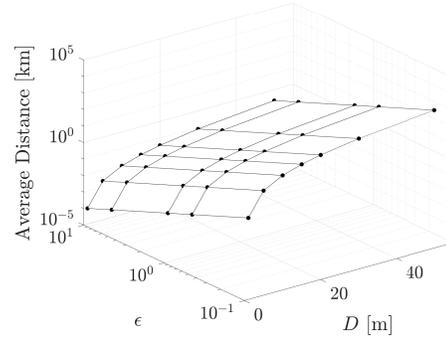}
    \caption{Average values of the \ac{AD} between the true and reported locations of the UAV, considering different values of $\epsilon$ and $D$ in the planar Laplacian distribution. }
    \label{fig:distance}
\end{figure}

We first evaluated the effect of \emph{DiPrID} to the degree of location privacy obtained by the UAV. To this aim, we considered different values of the parameters $\epsilon$ and $D$ in the planar Laplacian, and we evaluated the change they cause to the position reported by the UAV. In line with the literature (\cite{liu2021_tmc}), we evaluated the UAV location privacy in terms of \ac{AD}, i.e., the average 2-D distance (in terms of latitude and longitude) between the actual UAV location and the obfuscated one broadcasted in the messages. Fig.~\ref{fig:distance} reports the mean value of the \ac{AD} as a function of the specific parameters in the planar Laplacian.

Note that the \ac{AD} increases as $\epsilon$ decreases and $D$ increases. For instance, when $\epsilon=0.5$ and $D=5$~meters the \ac{AD} is $888.29$~meters, while it increases to $3,558$~meters considering the same value of $\epsilon$ and $D=10$~meters.
On the one hand, tuning appropriately the parameters $\epsilon$ and $D$ could provide enhanced location privacy properties to the \acp{UAV}. On the other hand, extreme values of such parameters would prevent the detection system from identifying collisions timely. Thus, the usage of only specific combinations of the parameters $\epsilon$ and $D$ should be allowed on the \acp{UAV}, so to allow the monitoring stations to detect unauthorized invasions.

\begin{figure}[htbp!]
    \centering
    \includegraphics[width=.8\columnwidth]{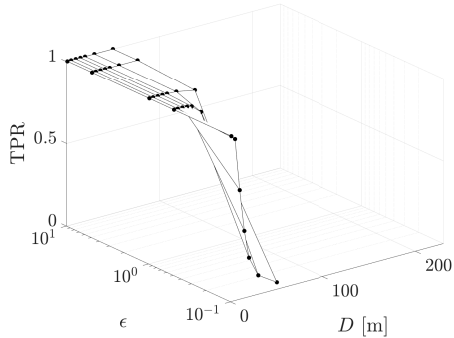}
    \caption{Average values of the \ac{TPR} of the detection ratio on all the receivers, for different values of $\epsilon$ and $D$ of the Planar Laplacian distribution used to perturb UAV position, assuming $W=15$~seconds.  }
    \label{fig:detection_eps_D_W5}
\end{figure}

To investigate the relationship between the parameters set on the \acp{UAV} and the CI detection capabilities, we set up some simulations where we evaluated the \ac{TPR} of the detection system while varying the parameters mentioned above, initially considering a single static value of $W$. Fig.~\ref{fig:detection_eps_D_W5} shows the \ac{TPR} of the detection system with different values of $\epsilon$ and $D$, considering $W=15$~s.
As previously discussed, extreme values of the parameters $\epsilon$ and $D$ make the detection system highly ineffective. For instance, when the $UAV$ adopts $\epsilon = 0.1$ and $D=30$~meters, the \ac{TPR} is only $0.114$, unacceptably low for a deployable detection system. However, choosing non-extreme values of the parameters mentioned above enhances the performance of the detection system. For instance, when $\epsilon = 0.5$ and $D=30$~meters, the \ac{TPR} increases to $0.942$.

\begin{figure}[htbp!]
    \centering
    \includegraphics[width=.8\columnwidth]{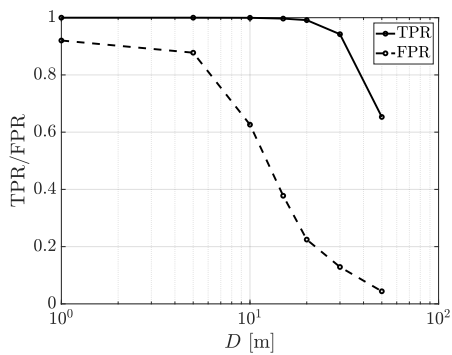}
    \caption{Average values of the \ac{TPR} and \ac{FPR} of the detection ratio on all the monitoring stations, for different values of the parameter $D$, assuming $\epsilon = 0.5$ and $W=15$~seconds. }
    \label{fig:detection_W_TPR_FPR}
\end{figure}

The price to pay for increased location privacy is the occurrence of false positives, i.e., erroneously detect an \emph{invading UAV}, even if no invasion was actually performed. To show this undesired phenomenon, in Fig.~\ref{fig:detection_W_TPR_FPR} we report both the \ac{TPR} and the \ac{FPR} of the invasion detection logic run by the monitoring station(s) of the CI operator, taking as a reference the scenario where $\epsilon=0.5$ and $W=15$~seconds, with different values of $D$.

As $D$ decreases (i.e., the reported locations are less noisy), the \ac{TPR} increase, but also the \ac{FPR} increase, as locations reported in the alert area are more likely to generate a detection. Indeed, when moving throughout the \emph{alert area}, one single \emph{invasion} event is necessary to declare the UAV as invading, leading to a false-positive. Interestingly, such an effect is compensated with high values of $D$, due to a higher level of noise affecting the location reports. The previously-mentioned configuration leads to a quicker decrease of the FPR compared to the decrease of the TPR. As previously highlighted, there are configurations leading to satisfactory results for both the UAV and the CI. Indeed, with $D=20$~m and $D=30$~m, the detection system achieves $TPR=0.991$ and $TPR=0.942$, as well as $FPR=0.225$ and $FPR=0.129$, respectively. Using such two configurations, as shown in the previous Fig.~\ref{fig:distance}, the UAV emits location at an average distance of $14.204$~m and $31.914$~km from the actual location, respectively. Clearly, in such conditions, the limiting factor for the location privacy of the UAV is the receiving range of the CI. Indeed, given that a message can be received wirelessly only if the actual distance between the transmitter (UAV) and the receiver (CI antenna) is less than $1,000$~km, the actual uncertainty on the true UAV location is exactly $1,000$~km (best-case privacy for the UAV, worst-case location uncertainty for the CI operator). As described in Sec.~\ref{sec:reporting}, the CI operator can identify and resolve false positives by interacting with the TTP, which is the only entity able to unveil the actual UAV location.

Finally, note that DiPrID likely introduces a detection delay, i.e., a time difference between the invasion of the no-fly zone and the detection of such event by the CI monitoring station. Indeed, for an invasion to be detected, one packet emitted by the UAV should be broadcasted with an added noise so as not to cause the resulting position to be outside of the no-fly zone. Assuming $W=15$~seconds, we computed an average detection delay of $1.13$~seconds, with $95$\% confidence interval of $\pm 1.1$~s. Worst-case, assuming a maximum UAV speed of $70$~km/h, the invading UAV could be detected after a maximum displacement of $\approx43$~m ($\approx 10$~\% of the monitored area), still leaving time to the operators to take countermeasures.

\section{Related Work}
\label{sec:related}

\acl{DP} has found applications in several scenarios, e.g., \acp{LBS}. For instance, the authors in~\cite{elsalamouny2016_tdp} proposed \ac{DP}-based probabilistic models for the users' locations, assuming the disclosure of multiple independent \ac{PoI}. The case of UAV Remote Identification is different, as \acp{UAV} share location information each $T \leq 1$~s and, hence, consecutive location data are not independent. In this case, Laplace-based \ac{DP} is demonstrated to be $\left(\epsilon, D \right)$-differentially private. The analysis in~\cite{elsalamouny2016_tdp} does not apply for consecutive location disclosures. The authors in~\cite{wang2019_fgcs} proposed to inject random noise into user-disclosed location and to combine it with the uncertainty provided by the inaccuracy of the GPS data. However, the scenario is different from that considered in our paper. The authors in~\cite{zhao2021_access} proposed a privacy-preserving trajectory data publishing method that can reduce global least-information loss and guarantee strong individual privacy. Their solution is effective, especially in the case of repetitive trajectory publishing. However, \acp{UAV} often have different mission-based trajectories. Thus, the proposed method does not apply to our scenario. The authors in~\cite{arif2021_meas} focused on the usage of vehicles locations/trajectories by \acp{LBS}, and leveraged k-anonymity and \ac{DP} to provide anonymity and trajectory privacy. 

In the context of the applications of \ac{DP} to \acp{UAV}, the authors in~\cite{kim2019_tvt} and~\cite{kim2017differential} proposed to optimize acp{UAV} movements to minimize privacy issues on users located in the traversed areas. Therefore, their work does not focus on  \acp{UAV} location privacy. Similar issues can be found in~\cite{fitwi2019_isc2}, which proposed a video processing solution that prevents \ac{UAV} from spying through windows. 

\ac{DP} has been applied also in Industrial Internet of Things (IIoT)~\cite{yin2017location}, crowdsensing~\cite{yang2018density}, crowdsourcing~\cite{wei2019differential}, edge computing~\cite{miao2019differential}, and big data~\cite{gu2018location}, all with different objectives.
Few contributions analyzed the trade-off between privacy preservation and data utility. The authors of~\cite{di2019catch} analyzed how geo-Ind impacts the utility when deploying WiFi hotspots for traffic offloading and for the spreading of information. The privacy-utility trade-off has been investigated also in~\cite{liu2021_tmc}, where the authors use geo-indistinguishable GPS traces to predict the users' travel time. Clearly, the scenarios and utility definitions in the above-cited works are different from ours.

Many studies investigated location privacy for \acp{VANET}, such as~\cite{kim2021_cose}. However, although the degrees of freedom in the movement of \acp{UAV} are higher than the ones of road-based transportation, smart vehicles in VANETs do not have to be constantly tracked for invasions of unauthorized areas. 

Overall, the previous contributions did not evaluate the application of \ac{DP}-related concepts to protect the \acp{UAV} location privacy, and neither considered location privacy issues connected with \emph{RemoteID}. Also, none of them evaluated the effects of such location protection on the effectiveness of area invasion detection systems, and neither the trade-offs between location privacy and data utility in this context. All the features mentioned above contribute to making our study novel and appealing for future research.

\section{Conclusion and Future Work}
\label{sec:conclusion}

In this paper, we investigated the trade-off between the location privacy that can be provided to \emph{RemoteID}-compliant drones and data utility, i.e., the capability of the detection systems deployed by \ac{CI} operators to timely detect invasions of \emph{no-fly zones}.

Through experimentation on real UAV data, we demonstrated that it is possible to obfuscate the location broadcasted by the UAV by a significant distance (e.g., $31.914$~km) while still allowing CI monitoring station(s) to detect invasions accurately, e.g, $94.2$\%. On the CI operator, such accuracy implies a slight detection delay ($1.13$~seconds), which can be easily considered when deploying the system. The other undesired effect is the presence of a few false positives, i.e., false invasions, solvable through interacting with the FAA, which is the only entity able to unveil the actual UAV location. Such location unveiling is only done if UAVs actually invaded the no-fly zone, preserving at most UAVs' location privacy.

Overall, our work demonstrates that it is possible to enhance UAVs' location privacy and still preserve data utility without hindering the future \ac{UAV} developments and applications.
Future work includes the extension of DiPrID with mechanisms to further enhance location privacy on UAVs, as well as detection techniques able to further reduce false positives.


\bibliographystyle{ACM-Reference-Format}

\bibliography{biblio}

\appendix
\section{Annex A: Notation}
\label{sec:annex}
We report the main notation used in our manuscript in Tab.~\ref{tab:notation}.

\begin{table*}[htbp]
\caption{Notation used throughout the paper.}
\centering
    \begin{tabular}{|p{2.3cm}|p{12.5cm}|}
    \hline
        \textbf{Not.} & \textbf{Description} \\ \hline
        $CI_i$ & Generic $i^{th}$ CI operator. \\ \hline
        $J_i$ & Number of monitoring stations (wireless receivers) deployed by the CI operator. \\ \hline
         $\delta_i$ & Safety radius of $CI_i$, i.e., radius of the no-fly zone. \\ \hline
         $R_{\rm i,max}$ & Theoretical receiving range of the monitoring station. \\ \hline
         $r_i$ & Amplitude of the circular crown defining the \emph{alert area}. \\ \hline
         $\mathcal{A}_1$, $\mathcal{A}_2$, $\mathcal{A}_3$ & Attackers. \\ \hline
         $\epsilon, D$ & Parameters of the planar Laplacian used to provide \ac{Geo-Ind} to \acp{UAV}. \\ \hline
         $x_{n,0}$ & Actual non perturbed location of the UAV $n$. \\ \hline
         $\theta$ & Angle that the segment joining $x$ and $x_{n,0}$ forms with the horizontal axis. \\ \hline
         $F_{\epsilon}(D,\theta)$ & Equation of the planar Laplace distribution, as a function of the parameters $\epsilon$, $D$, and $\Theta$. \\ \hline
         $\gamma_{-1}$ & $-1$ branch of the Lambert function, used to compute $D$. \\ \hline
         $p$ & Parameter extracted from a uniform distribution in $\left[ 0,1 \right]$, used to compute $D$. \\ \hline
         $x_{n,0}^m$ & Vector containing the location information of the UAV $n$ at time instant $m$. \\ \hline
         $x_n^m$ & Location broadcasted by the UAV $n$ at the time $m$, obfuscated with \ac{Geo-Ind}. \\ \hline
         $\mathcal{L}$ & Set of admissible locations independent from the real location $x_{n,0}$. \\ \hline
         $W$ & Duration of the \emph{Detection Window} on the \ac{CI} monitoring station(s). \\ \hline
         $geo\_pos_i$ & Location of the $i$-th receiver in geographical coordinates. \\ \hline
         $xyz\_pos_i$ & Location of the $i$-th receiver in 3-D Cartesian coordinates. \\ \hline
         $W_{en}$ & Flag indicating if the evaluation of $W$ already started. \\ \hline
         $msg_m$ & Generic \emph{RemoteID} message. \\ \hline
         $ID_m$ & (Pseudo)-Identity of the UAV broadcasting the $m$-th \emph{RemoteID} message. \\ \hline
         $t_m$ & Timestamp in the $m$-th \emph{RemoteID} message. \\ \hline 
         $lat_m$, $lon_m$, $alt_m$ & Latitude, longitude, and altitude reported in the \emph{RemoteID} message.\\ \hline
         $v\_{x_m}$, $v\_{y_m}$, $v\_{z_m}$ & Speed on $x$, $y$, and $z$ axis reported in the \emph{RemoteID} message.\\ \hline 
         $d_{i,m}$ & Distance computed on the $i$-th monitoring station from the UAV emitting $msg_m$. \\ \hline
         $\theta_m$ & Angle between the location estimated by the $i$-th monitoring station and the UAV emitting $msg_m$. \\ \hline
         $\mathbf{det\_events}$ & Vector of detection events ($\{ 0,1 \}$ within $W$. \\ \hline
         $\tau$ & Detection delay on the CI monitoring station. \\ \hline
         $p_{TTP}, P_{TTP}$ & Private and public key of the TTP. \\ \hline
         $c_{n,0}^m$ & Encrypted location report generated by the UAV $n$ for the message $m$. \\ \hline
    \end{tabular}
    \label{tab:notation}
\end{table*}

\end{document}